\documentclass[aps,prl,twocolumn]{revtex4}

\usepackage{graphicx}

\usepackage{amsfonts}
\usepackage{amssymb}
\usepackage{amsmath}
\usepackage{braket}

\newcommand{\dv}[1]{\,\mathrm{d}#1\,}
\newcommand{\qop}{\mathit{O}}
\newcommand{\fq}[1][\rho,\qop]{F_\text{Q}\left[#1\right]}
\newcommand{\Cop}{\mathit{C}}
\newcommand{\cop}{\mathit{c}}
\newcommand{\var}[1]{\text{Var}\,#1}
\newcommand{\qpar}{q\,}

\usepackage{natbib}

\usepackage[capitalize]{cleveref}

\begin{document}
\title{From entanglement certification with quench dynamics\\ to multipartite entanglement of interacting fermions}
\author{Ricardo \surname{Costa de Almeida}}
\author{Philipp \surname{Hauke}}  

\affiliation{INO-CNR BEC Center and Department of Physics, University of Trento, Via Sommarive 14, I-38123 Trento, Italy}

\affiliation{Kirchhoff Institute for Physics, Ruprecht Karl University of Heidelberg, Im Neuenheimer Feld 227, D-69120 Heidelberg, Germany}

\affiliation{Institute for Theoretical Physics, Ruprecht Karl University of Heidelberg, Philosophenweg 16, D-69120 Heidelberg, Germany}

\date{\today}
\begin{abstract}
	Multipartite entanglement, such as witnessed through the quantum Fisher information (QFI), is a crucial resource for quantum technologies, but its experimental certification is highly challenging.
	Here, we propose an experimentally friendly protocol to measure the QFI. 
	It relies on recording the short-time dynamics of simple observables after a quench from a thermal state, works for spins, bosons, and fermions, and can be implemented in standard cold-atom experiments and other platforms with temporal control over the system Hamiltonian.
	To showcase the protocol, we  simulate it for the one-dimensional Fermi--Hubbard model. 
	Further, we establish a family of bounds connecting the QFI to multipartite mode entanglement for fermionic systems, which enable the detection of multipartite entanglement at sizable temperatures. 
	Our work paves a way to experimentally accessing entanglement for quantum enhanced metrology.	
\end{abstract}
\maketitle
\paragraph{\textbf{Introduction.}}
A central question for quantum many-body physics is to understand the structure of entanglement and how it translates into observable features.
Besides its potential to explain certain salient many-body phenomena \cite{AmicoFazioVedral2008,Laflorencie2016,DeChiaraSanpera2018,AbaninBlochSerbyn2019,GogolinEisert2016,Wen2017}, it may take a decisive role as a resource in upcoming quantum technologies.
Hence, as these technologies mature, scalable protocols for detecting entanglement become increasingly necessary \cite{GuhneToth2009,FriisHuber2018}.
This demand is already a reality for quantum metrology \cite{PezzeSmerziOberthalerTreutlein2018} where the quantum Fisher information (QFI) \cite{BraunsteinCaves1994}, a witness for multipartite entanglement, determines the metrological quantum enhancement \cite{PezzeSmerzi2009,Toth2012,HyllusPezzeSmerzi2012,Toth2014}.
Although lower bounds of the QFI have been obtained in recent groundbreaking experiments \cite{StrobelPezzeSmerziOberthaler2014,LuckeKlempt2011,LuckeKlempt2014,BohnetBollinger2016}, general and efficient procedures to directly extract its precise value in many-body systems are lacking.

To tackle this challenge, we develop an experimentally accessible technique for measuring the QFI for states in thermal equilibrium.
In contrast to a previous proposal relying on frequency-dependent dynamic susceptibilities \cite{HaukeHeylTagliacozzoZoller2016}, our protocol only requires measuring the short-time dynamics of mean expectation values after a quench.
This straightforward procedure is ideally suited, e.g., for standard experiments on ultra-cold  atoms \cite{Bloch2005,LewensteinSanperaSen2007,HaukeTagliacozzoLewenstein2012,EisertGogolin2015,LangenSchmiedmayer2015}.  
This measurement protocol for the QFI is our first main result.  

\begin{figure}
	\includegraphics[width=0.9\columnwidth]{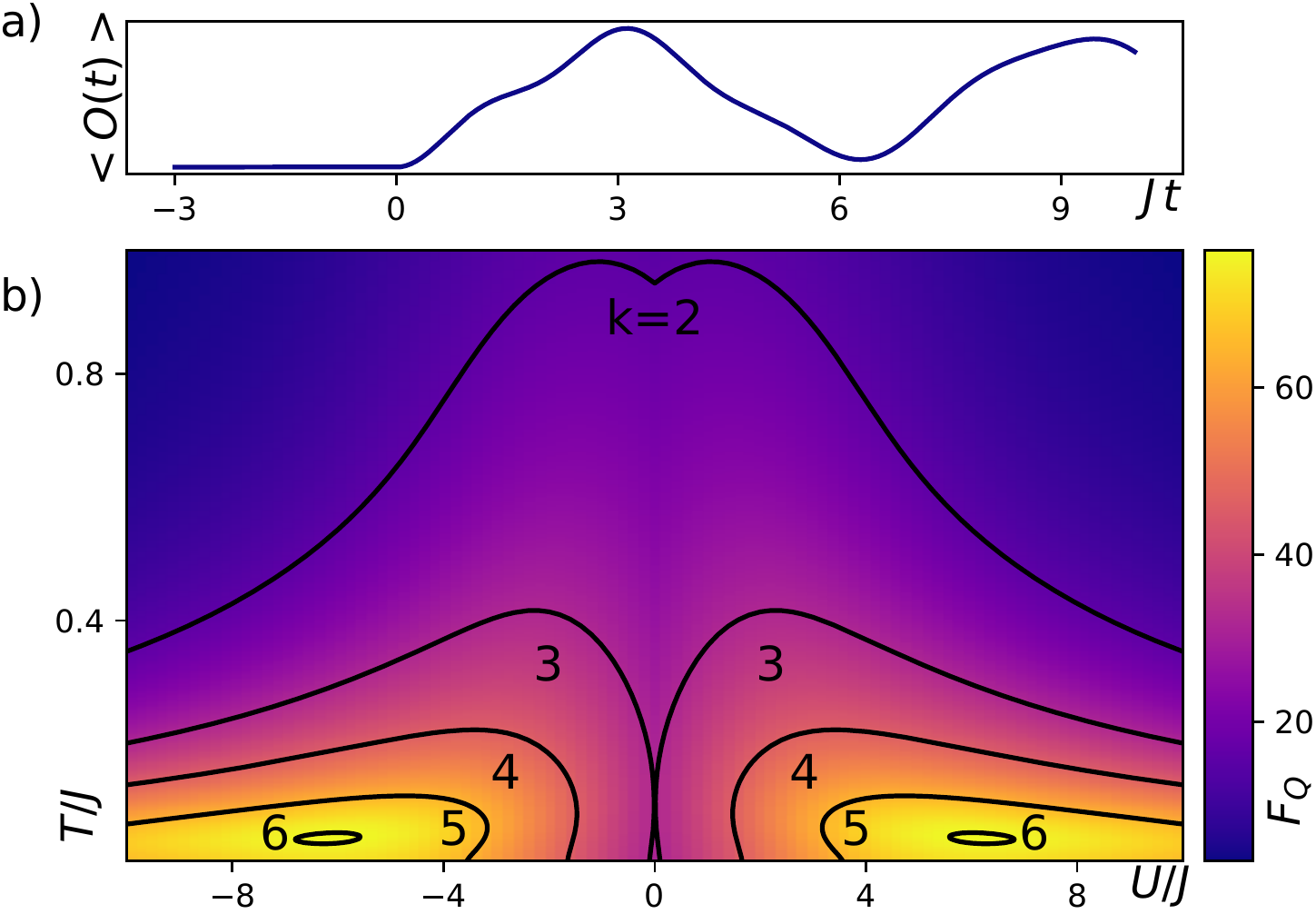}
	\caption{\textbf{Certification of multipartite mode entanglement.} 
		\textbf{(a)} A many-body quantum system is abruptly perturbed by an operator \(\qop\).
		Subsequently, the evolution of the same observable \(\braket{\qop(t)}\) is measured, from which the quantum Fisher information, \(\fq\), is extracted.
		\textbf{(b)} \(\fq\), computed by simulating the protocol for the Fermi--Hubbard model in 1D.
		Using the entanglement bounds of \cref{ineq:qfi:kprodspecific} (contour lines), many-body entanglement is certified up to large temperatures.
		Data for \(L=8\) and open boundary conditions.
		}
	\label{fig:main}
\end{figure}

Moreover, previous studies about the QFI and entanglement bounds have focused on systems describable as spins \cite{PezzeSmerzi2009,Toth2012,HyllusPezzeSmerzi2012}.
Nonetheless, interacting fermions are of central importance to condensed matter physics and experiments with ultra-cold atoms have enabled the precise engineering of fermionic many-body systems \cite{GiorginiPitaevskiiStringari2008,Esslinger2010,SerwaneJochim2011,ParsonsGreiner2015,MurmannBergschneiderJochim2015,ChiuGreiner2018,TarruellSanchezPalencia2018}.

Motivated by this, we derive bounds that relate multipartite fermionic mode entanglement to the QFI, by generalizing the concept of \(k\)-producibility to fermionic systems. 
This framework for fermionic multipartite entanglement is our second main result. 

We illustrate these bounds as well as our quench-based measurement protocol for the QFI at a paradigmatic example, the Fermi--Hubbard model in one dimension (1D).
As shown in \cref{fig:main} and discussed further below, we certify the presence of multipartite mode entanglement for a broad region of the parameter space. 

The article is organized as follows: First, we review some basic notions regarding the QFI.
We proceed to derive our quench protocol.
Afterwards, we rigorously define multipartite mode entanglement for fermions and determine the correct fermionic entanglement bounds for the QFI.
Subsequently, we discuss the results shown in \cref{fig:main} in detail and provide a guideline to experiments aiming at certifying entanglement in the Fermi--Hubbard model.
We conclude the article with a brief outlook. 

\paragraph{\textbf{Background on the QFI.}}
The quantum Fisher information, \(\fq\), is a central concept in quantum metrology. 
It quantifies the metrological sensitivity obtained from a given quantum state \(\rho\) in a phase estimation setup, in which a unitary generated by an operator \(\qop\) rotates \(\rho\) by an angle \(\theta\), \(\rho(\theta) = e^{-i\theta \qop}\rho e^{i\theta\qop}\) \cite{PezzeSmerzi2014}. 
The aim in this scenario is to precisely estimate the parameter \(\theta\), whose variance after \(m\) measurements is bounded through the Cramér--Rao bound, \( \var{\theta} \geq 1/\left(m \fq \right)\) \cite{BraunsteinCaves1994}.

Moreover, the QFI witnesses multipartite entanglement.
Specifically,  we show this below for fermionic states \(\rho\) defined by \(|M|=dk+r\) fermionic modes.
If  \(\rho\) satisfies
\begin{align}
	\fq > \left(dk^2+r^2\right)\, \text{,}
	\label{ineq:qfi:kprodspecific}
\end{align}
it must be, at least, \((k+1)\)-partite mode entangled.
This result complements existing, analogous bounds for spin systems \cite{PezzeSmerzi2009,Toth2012,HyllusPezzeSmerzi2012}.
Intuitively, the higher correlations of an entangled many-body state,  relative to a classical state, lead to a greater sensitivity to perturbations and thus to greater metrological gain, quantified through \(\fq\).
In particular, in a separable state \(\fq\) is bounded by \(|M|\), the scaling observed in classical systems, whereas any metrological enhancement beyond the classical limit requires entanglement.
This enables the use of \(\fq\) for entanglement certification.

The QFI of a pure state \(\rho = \ket{\psi}\bra{\psi}\) is simply a variance, \(\fq =4\var{\qop} =4\left(\bra{\psi}\qop^2\ket{\psi}-\bra{\psi}\qop\ket{\psi}^2\right)\), so it can be calculated efficiently. However, the formula for an arbitrary density matrix \(\rho = \sum_\lambda \rho_\lambda \ket{\lambda}\bra{\lambda}\), 
\begin{align}
	\fq= 2\sum_{\lambda , \lambda'} \frac{\rho_\lambda-\rho_{\lambda'}}{\rho_\lambda+\rho_{\lambda'}}\left( \rho_\lambda-\rho_{\lambda'} \right)|\bra{\lambda}\qop\ket{\lambda'}|^2\,\text{,}
	\label{def:qfi:eigenbasis}
\end{align}
requires diagonalizing the state, which is a challenging undertaking for quantum many-body systems, both theoretically and experimentally. In what follows, we show how to circumvent this difficulty by extracting the QFI for thermal states from expectation values using a quench. 

\paragraph{\textbf{Derivation of the quench protocol.}}
In a previous work \cite{HaukeHeylTagliacozzoZoller2016}, a connection between the QFI and linear response theory  was found that enables one to compute the QFI for systems in equilibrium at temperature \(T\),
\begin{align}
	\fq = \frac{4}{\pi}\int_0^{+\infty} d \omega \tanh\left(\frac{\omega}{2T}\right)\chi"(\omega,T)\,\text{.}
	\label{eq:qfi:chiw}
\end{align}
This formula requires knowledge of \(\chi"(\omega,T)=\Im(\chi(\omega,T))\), the imaginary part of the Fourier transform of the response function
\begin{align}
	\chi(t-\tau,T) = \theta(t-\tau) \braket{[\qop(t),\qop(\tau)]}\,\text{.}
	\label{def:lresp:chi}
\end{align}
This function characterizes the linear response of an observable \(\braket{\qop(t)}\) to a time-dependent perturbation from \(H_0\) to \(H(t)= H_0-f(t)\qop\), where \(H_0\) is the Hamiltonian with respect to which the system was at thermal equilibrium.
For deviations \(\Delta\qop(t)\) from the equilibrium value, the Kubo formula gives \cite{Kubo1966}
\begin{align}
	\Delta \qop(t) &=\braket{\qop(t)}-\braket{\qop}= \left(\chi * f\right)(t) \notag\\
		       &= \int_{-\infty}^{+\infty} \dv{\tau} \chi(t-\tau,T)f(\tau)\,\text{.}
	\label{eq:lresp:kubo}
\end{align}

By transforming the integral from frequency to time domain, an equation analogous to \cref{eq:qfi:chiw} follows 
\begin{align}
	\fq = 4 T\int_0^{+\infty} \dv{t} \frac{\chi(t,T)}{\sinh\left(\pi t T\right)}\,\text{,}
	\label{eq:qfi:chit}
\end{align}
which allows the QFI to be obtained directly from the Kubo response function.
The time domain expression has computational advantages compared to \cref{eq:qfi:chiw} and has been used for computing the QFI \cite{HaukeHeylTagliacozzoZoller2016,GabbrielliSmerziPezze2018}. 

Conceptually, these expressions represent a significant advance as they explicitly relate the QFI to correlations encoded in the response functions.
However, their application still presents practical problems as measurements of unequal-time correlation functions, such as \(\braket{[\qop(t),\qop]}\), are often challenging. 

We overcome such limitations by introducing a protocol that solely relies on measurements of expectation values \(\braket{\qop(t)}\).
To realize such a simplified protocol only requires a weak, abrupt quench, as can be conveniently implemented, e.g., in cold-atom  experiments \cite{LewensteinSanperaSen2007,HaukeTagliacozzoLewenstein2012,EisertGogolin2015,LangenSchmiedmayer2015}.
In this scenario, the drive function is simply \(f(\tau)=\qpar \theta(\tau)\), so, from \cref{eq:lresp:kubo}, the dynamics are governed by 
\begin{align}
	\Delta \qop(t)_{\text{quench}}=\qpar\int_0^t \dv{\tau} \chi(\tau,T) = \qpar\xi(t,T) \,.
	\label{eq:lresp:quench}
\end{align}
Here, \(\qpar\) denotes the quench amplitude and we introduced \(\xi(t,T)=\Delta \qop(t)_{\text{quench}}/\qpar\).
Using \(\chi(t,T)=\dv{\xi(t,T)}/\dv{t}\) in \cref{eq:qfi:chit}, we arrive at
\begin{align}
	\fq=\frac{4\pi T^2}{\qpar}\int_0^{+\infty} \dv{t} \frac{\Delta \qop(t)_\text{quench}}{\sinh\left(\pi t T\right)\tanh\left(\pi t T\right)}
	\label{eq:qfi:quench}
\end{align}
after performing an integration by parts and handling the convergence issues that arise. See the supplementary material for details.  

\begin{figure}
	\includegraphics[width=\columnwidth]{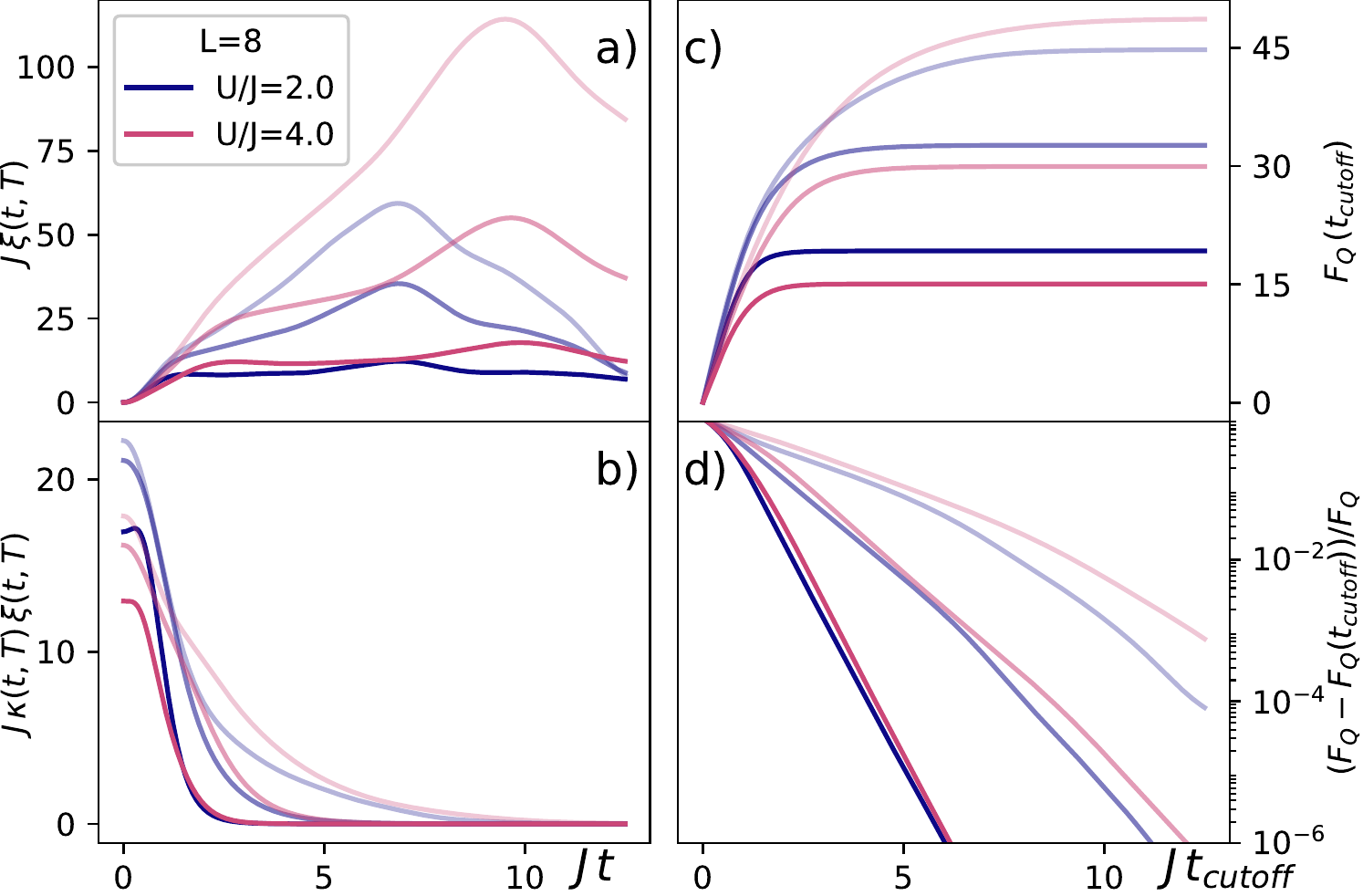}
	\caption{\textbf{Quench protocol for QFI extraction,} exemplified for the Fermi--Hubbard model at temperatures \(T/J=0.2,\,0.4,\,0.8\) (from light to dark shades) for a quench with the staggered magnetization (\cref{def:op:staggered}).
		\textbf{(a)} At time \(t=0\), the system is quenched with the operator \(\qop\) and strength \(q\).
		Measuring the deviations from the equilibrium expectation value yields \(\xi(t,T) = \Delta \qop(t)_{\text{quench}}/q\).
		\textbf{(b)} Using  \cref{eq:qfi:quench}, the QFI can be computed by integrating \(\xi(t,T)\) multiplied with the kernel function, \(\kappa(t,T)\).
		\textbf{(c)} Cutting the integral off at time \(t_\text{cutoff}\) produces a lower bound \(\fq(t_\text{cutoff})\le\fq\).
		\textbf{(d)} Due to the functional form of \(\kappa(t,T)\), the convergence is exponentially fast with a decay constant set by the temperature, \(\fq-\fq(t_\text{cutoff})\sim \exp(-\pi T t_\text{cutoff})\).
		}
	\label{fig:protocol}
\end{figure}

From \cref{eq:qfi:quench}, we can summarize our protocol  by four steps (see also \cref{fig:protocol}):
\textit{(i)} Prepare a thermal state.
\textit{(ii)} Turn on quench.
\textit{(iii)} Measure dynamics of expectation values.
\textit{(iv)} Integrate results according to \cref{eq:qfi:quench}. 

Thermal equilibrium and a quench in the linear regime are the only assumptions used for deriving \cref{eq:qfi:quench}, so the protocol applies to arbitrary quench operators and quantum many-body systems, including fermionic, bosonic, and spin systems.
Moreover, it has a series of  advantageous properties.
For example, it simplifies the requirements for extracting the QFI in many situations, as no time--time correlations are required, and the exponential decrease of \(\kappa(t,T)=4\pi T^2 \left[\sinh\left(\pi t T\right)\tanh\left(\pi t T\right)\right]^{-1}\) with time implies only short measurement times are required. 

\paragraph{\textbf{Entanglement bounds for fermionic systems.}} Based on the concept of \(k\)-producibility, for spin systems bounds on the QFI have been derived that only states with multipartite entanglement can overcome \cite{PezzeSmerzi2009,Toth2012,HyllusPezzeSmerzi2012}.
However, for fermionic systems, such bounds do not exist. 

To remedy this situation, we first need to adapt the notion of \(k\)-producibility.
To see why this is necessary, we recall the condition for a state of \(N\) spins to be \(k\)-producible,
\begin{align}
	\ket{\psi}_{k\text{-prod.}}^\mathrm{spin}=\ket{\psi_1}\otimes\ket{\psi_2}\otimes \dots\otimes\ket{\psi_{P-1}}\otimes\ket{\psi_P}\,,
	\label{def:state:kprodspin}
\end{align}
where \(\ket{\psi_j}\) is a state of \(N_j\leq k\) spins and \(\sum_j N_j =N\). 
Such a decomposition is not meaningful in the fermionic case due to the antisymmetric structure of the wave function.
Fortunately, this also suggests what is the correct criteria, which we now introduce.

Consider a set of fermionic modes \(M\), with associated creation and annihilation operators \(\cop^\dagger_m\) and \(\cop_m\), labeled by \(m\in M\).
A \(k\)-partition of the system is defined as a partition \(M= M_1\cup M_2 \cup\dots \cup M_P\) subject to \(|M_j|\leq k\).
Now, we introduce the following definition: a pure fermionic state \(\ket{\psi}\) is \(k\)-producible if there is a \(k\)-partition such that
\begin{align}
	\begin{split}
		\ket{\psi}_{k\text{-prod.}}&=\Cop^\star_1\Cop^\star_2\dots\Cop^\star_P\ket{}\\
	\end{split}\,,
	\label{def:state:kprodfermion}
\end{align}
where the operator \(\Cop^\star_j\) is restricted to act on \(M_j\).
The \(\Cop^\star_j\) can be written as linear combinations of products of creation operators $\cop_m^\dagger$ acting within \(M_j\),  
\begin{align}
	\Cop_j^\star = \sum_{\eta_j} \phi_j^\star(\eta_j)\prod_{m\in M_j} \left(\cop_m^\dagger\right)^{\eta_j(m)}\, \text{.} 
	\label{def:op:kprodfermion}
\end{align}
Here, without loss of generality, we fix some order for applying the creation operators. 
The summands are labeled by numbers \(\eta_j(m)\in\{0,1\}\), which one can envision as the possible occupations of the modes, with associated amplitudes \(\phi^\star_j(\eta_j)\in\mathbb{C}\).

An explicit connection with the spin definition is possible if we introduce \(\ket{\psi_j}=\Cop^\star_j\ket{}\) and notice that \cref{def:state:kprodfermion} can be written as \(\ket{\psi}\sim\ket{\psi_1}\wedge\ket{\psi_2}\wedge\dots\wedge\ket{\psi_P}\), with the exterior product \(\wedge\) acting as an antisymmetric analogue of the tensor product.
The \(1\)-producible decomposition with the exterior product has been used before to study mode entanglement in fermionic systems \cite{FriisBruschi2013}.
Nonetheless, for our purposes the operator language as in \cref{def:op:kprodfermion} is more convenient.
The same formulation can be adapted to bosonic and spin systems, where it reproduces the usual definition of \(k\)-producible states.
For \(2\)-partite entanglement, our definition is equivalent to the one through the Slater number \cite{SchliemannCiracLewensteinLoss2001,EckertLewenstein2002}.

The extension of these concepts to mixed states \(\rho\) is standard \cite{PezzeSmerzi2009,Toth2012,HyllusPezzeSmerzi2012}: a mixed state \(\rho_{k\text{-sep.}}\) is \(k\)-separable if it can be written as a convex hull
\begin{align}
	\rho_{k\text{-sep.}}=\sum_\lambda \rho_\lambda \ket{\lambda}_{k\text{-prod.}}\bra{\lambda}_{k\text{-prod.}}
	\label{def:state:ksepfermion}
\end{align}
of \(k\)-producible states \(\ket{\lambda}_{k\text{-prod.}}\).
This formulation introduces a hierarchy for mixed states that defines multipartite entanglement of fermionic modes: a state is \((k+1)\)-partite mode entangled if it is not \(k\)-separable.

Using this notion, we can now establish bounds on multipartite mode entanglement.
To connect to the QFI, we focus on operators of the form 
\begin{align}
	\qop = \sum_{m\in M} w(m) \cop^\dagger_m \cop_m
	\label{def:op:bounds}
\end{align}
with \(w(m)\in\mathbb{R}\) weighting the occupation of different modes.
Given a \(k\)-producible state \(\rho=\ket{\psi}_{k\text{-prod.}}\bra{\psi}_{k\text{-prod.}}\), one can define a probability distributions \(p_j(\eta_j)=|\phi_j(\eta_j)|^2\) for the \(\eta_j\) and associated random variables \(w_j(\eta_j)=\sum_{m\in M_j} w(m)\eta_j(m)\) such that \(\fq=4\sum_j \var{w_j}\).
Employing Popoviciu's inequality \cite{Popoviciu1935} to bound \(\var{w_j}\), it follows that 
\begin{align}
	\fq\leq 4\sum_j\frac{1}{4} \left(\max_{\eta_j}w_j(\eta_j)-\min_{\eta_j}w_j(\eta_j)\right)^2\,.
	\label{ineq:qfi:popoviciu}
\end{align}
Additional knowledge about the state \(\ket{\psi}_{k\text{-prod.}}\) leads to restrictions on the allowed occupations \(\eta_j\) and permits the derivation of tighter bounds for \cref{ineq:qfi:popoviciu}.
In particular, if \(\ket{\psi}_{k\text{-prod.}}\) has a fixed occupation number 
\begin{align}
	\fq \leq \frac{d k^2+r^2}{4} \left(\max_m w(m)-\min_m w(m)\right)^2 \,,
	\label{ineq:qfi:kprod}
\end{align}
where we used the decomposition \(|M|=dk+r\). 
See the supplementary material for a detailed discussion and tighter bounds for the case where the occupation number is known.

Equation \cref{ineq:qfi:kprod} and related bounds immediately extend to \(k\)-separable mixed states due to the convexity of \(\fq{}{}\). As a consequence, any state that overcomes this bound  cannot be \(k\)-separable and must be, at least, \((k+1)\)-partite mode entangled. 

\paragraph{\textbf{Results for Fermi--Hubbard chain.}}
We illustrate our main results on the 1D Fermi--Hubbard model, a paradigmatic model for an interacting, fermionic many-body system.
Its Hamiltonian reads
\begin{align*}
	H_0 = -J\sum_{x,\sigma} \left(\cop^\dagger_{\sigma x}\cop_{\sigma x+1}+ h.c.\right) + U\sum_x \left(\cop^\dagger_{\downarrow x}\cop_{\downarrow x}\cop^\dagger_{\uparrow x}\cop_{\uparrow x} \right)\,.
\end{align*}
The fermions live on lattice sites \(x=1,2,\dots L\) and have two internal states, \(\sigma=\uparrow,\downarrow\).
\(J\)  governs hopping between neighboring sites and \(U\) controls on-site interactions.
The Hamiltonian \(H_0\) commutes with total occupation and magnetization, and we choose to work on the magnetization-free subspace at half-filling.   

To evaluate the QFI via \cref{eq:qfi:quench}, we consider quenches using the staggered magnetization, \(\qop_+\), and density, \(\qop_-\),
\begin{align}
	\qop_{\pm} = \sum_{x} (-1)^x \left(\cop^\dagger_{\uparrow x}\cop_{\uparrow x} \mp\cop^\dagger_{\downarrow x}\cop_{\downarrow x}\right) \, .
	\label{def:op:staggered}
\end{align}
This choice is motivated by limit cases: at \(U\rightarrow +\infty\) and half-filling, the fermions form a Néel state with homogeneous density and alternating internal state.
Here, \(\qop_+\) differentiates between the two degenerate ground states describing two possible alternating orders.
For \(U\rightarrow -\infty\), the fermions pair up to form a charge-density wave with homogeneous magnetization.
Here, \(\qop_-\) distinguishes two possibilities of alternating large and low density.
These limiting situations can be described analytically by an effective antiferromagnetic theory \cite{TarruellSanchezPalencia2018}.
Based on intuition from previous work \cite{HaukeHeylTagliacozzoZoller2016}, we expect \(\qop_\pm\) to give a large QFI as one goes from the free theory at \(U/J\to 0\) to the antiferromagnetic limit.

\begin{figure}
	\includegraphics[width=\columnwidth]{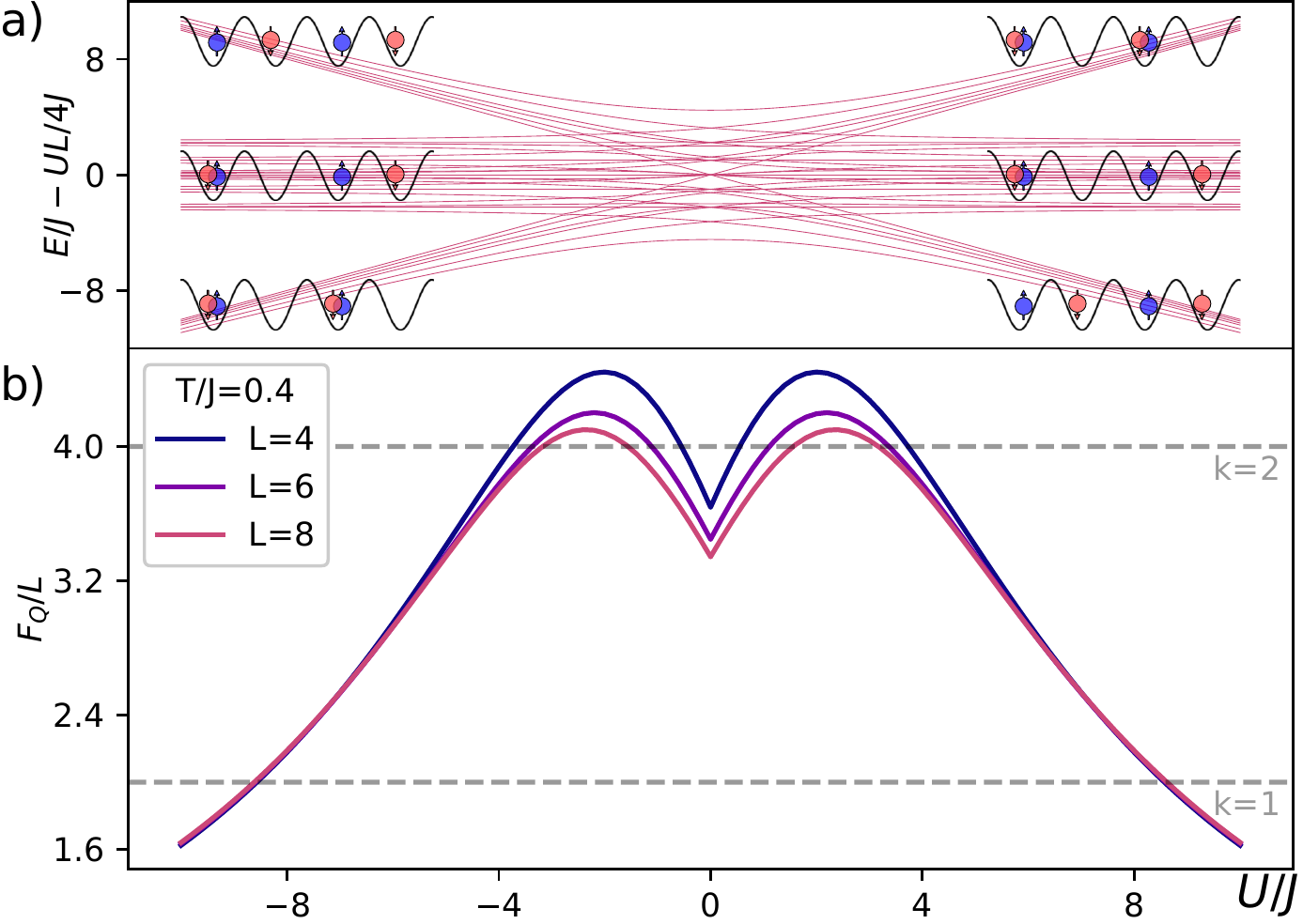}
	\caption{\textbf{Signature of robust entanglement in a Fermi--Hubbard 1D chain.}
		\textbf{(a)} Spectrum of the model, exemplified for \(L=4\).
		At \(U/J\to 0\), the system is a free theory with regular level spacings.
		The bands for \(U/J\to -\infty\) (\(U/J\to\infty\)) are a direct signature of the effective antiferromagnetic description.
		Higher bands correspond to breaking of pairs (creation of doublon-holon excitations).
		This picture falls apart for intermediate \(U\) as the the system moves out of the perturbative regime and displays strongly correlated behaviour.
		\textbf{(b)} \(\fq{}{}\) density for different system sizes with thresholds for certifying entanglement(dotted lines).
		The breakdown of the effective theory coincides with an increase in the robustness of the entanglement certified against thermal effects.
		}
	\label{fig:hubbard}
\end{figure}

To simulate the quench protocol, we extract \(\xi(t,T)\) from exact diagonalization and use it to calculate \(\fq\), taking the larger one of \(\fq[\rho,\qop_+]\) and \(\fq[\rho,\qop_-]\).
The results are summarized in Figs.~\ref{fig:main} and \ref{fig:hubbard}.
\(\fq\) increases rapidly as one moves away from the non-interacting point.
In particular, in the intermediate region, where neither the free nor the antiferromagnetic theory describes the system, multipartite entanglement is detected at temperatures as large as \(T/J= 0.4\).
The system-size dependence suggests the entanglement to be especially robust in this strongly interacting region, making it a prime candidate to search for experimental signatures of multipartite entanglement.

\begin{figure}
	\includegraphics[width=\columnwidth]{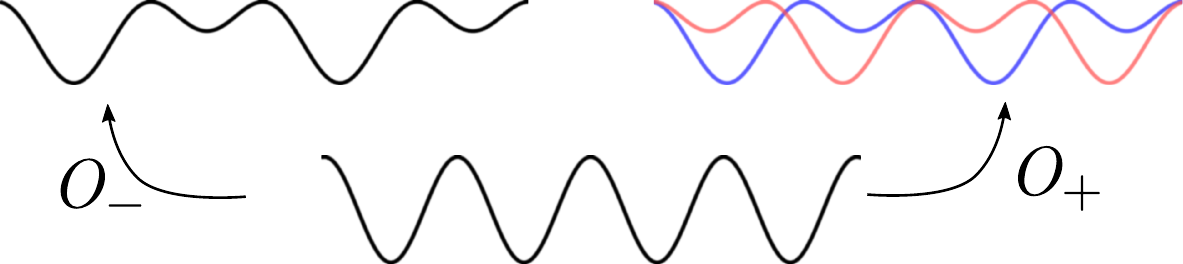}
	\caption{\textbf{Lattice quench in the Fermi--Hubbard model.} A quench with \(\qop_\pm\) amounts to abruptly modifying the chemical potential in a staggered fashion, which can be simply implemented through superlattices, which are spin-dependent for \(\qop_+\), without the need for quantum gas microscopes.
		The relevant observable \(\braket{\qop_\pm(t)}\) can be measured through site-dependent imaging \cite{YangOttZacheHalimehHaukePan2020}.
		}
	\label{fig:experimental}
\end{figure}

Figure~\ref{fig:experimental} illustrates how one can straightforwardly realize the quenches with \(\qop_-\) and \(\qop_+\) using optical superlattices.
Ultracold atoms are now reaching strongly-correlated many-body states of the Fermi--Hubbard model at temperatures as low as \(T/J=0.25\) \cite{MazurenkoGreiner2017,SalomonGross2018,VijayanBlochGross2020}, well within the region where multipartite entanglement can be detected (see \cref{fig:main}). 
Moreover, as shown in  \cref{fig:protocol}c,d, at such temperatures the QFI converges within few hopping events (\(Jt\lesssim 8\)), i.e., on time scales faster than typical decoherence rates \cite{VijayanBlochGross2020}.
Thus, our quench protocol enables the detection of multipartite entanglement within existing experimental setups. 

\paragraph{\textbf{Conclusion.}}
Though discussed in the context of ultracold fermionic gases, the simplicity and generality of our protocol make it readily applicable across different platforms.
It is also straightforward to replace the simple quench we have chosen by other time-dependent functions \(f(t)\), which just requires modifying the kernel function \(\kappa(t,T)\)(See the supplementary material for details). 
Recent works have studied the dynamical behavior of the quantum Fisher information after a quantum quench \cite{SmithHaukeHeylMonroe2016,PappalardiFazio2017}.
Here, turning things on their head, we have demonstrated the power of induced dynamics to extract the quantum Fisher information. 
Beyond the setup developed here, there is the possibility of applying our protocol to different thermodynamical ensembles \cite{BrenesPappalardiGooldSilva2020} and even extend it outside the realm of thermodynamical states \cite{MehboudiSanperaParrondo2018}.

\begin{acknowledgments}
	\paragraph{\textbf{Acknowledgments.}}
	This work is part of and supported by the DFG Collaborative Research Centre "SFB 1225 (ISOQUANT)", the Provincia Autonoma di Trento and the ERC Starting Grant StrEnQTh (Project-ID  804305). 
\end{acknowledgments}

\bibliography{references}

\onecolumngrid
\appendix*

\renewcommand{\theequation}{S\arabic{equation}}
\renewcommand{\thefigure}{S\arabic{figure}}
\renewcommand{\bibnumfmt}[1]{[S#1]}
\vspace*{1cm}
        
\begin{center}
	{{\Large{\textbf{Supplementary Material}}}}
\end{center}
       
\vspace*{0.25cm}
        
In this supplementary material, we give details on the mathematical derivations of the quench protocol, including the general expression for arbitrary driving functions \(f(t)\) as well as numerical studies of the convergence with the quench strength \(\qpar\).
Further, we provide additional details on the derivations of the entanglement bounds as well as refined expressions for a fixed particle number.

\section{Quench Protocol}

In this section, we give further details on the derivation of the detection scheme, in particular considering for arbitrary quench protocols.
Moreover, we discuss experimental issues, such as a noise, finite ramping times, and non-infinitesimal quench amplitudes.

\subsection{Details on the derivation of the detection scheme for arbitrary quench protocol}

To derive \cref{eq:qfi:chit}, one first notices that, in the canonical ensemble at temperature \(T\),
\begin{align*}
	\frac{\rho_\lambda-\rho_{\lambda'}}{\rho_\lambda+\rho_{\lambda'}} = \tanh\left(\frac{\omega_{\lambda\lambda'}}{2T} \right) = iT \int_{-\infty}^{\infty}\dv{t}\frac{e^{-i\omega_{\lambda\lambda'} t}}{\sinh\left(\pi t T\right)}   \text{, }
\end{align*}
where \(\omega_{\lambda\lambda'}=\epsilon_{\lambda'}-\epsilon_\lambda\) denotes the energy splitting between the two energy levels \(\ket{\lambda}\) and \(\ket{\lambda'}\).
Combining this expression with \cref{def:qfi:eigenbasis} and inserting a Heaviside step function \(\theta\left(t\right)\) yields 
\begin{align*}
	\fq & = 2\sum_{\lambda,\lambda'}  \tanh\left(\frac{\omega_{\lambda\lambda'}}{2T}  \right) \left(\rho_\lambda-\rho_{\lambda'}\right) |\qop_{\lambda \lambda'}|^2 = 4T\int_{0}^{\infty}\dv{t}\frac{1}{\sinh\left(\pi t T\right)}\left( i\theta(t) \sum_{\lambda,\lambda'}e^{-i\omega_{\lambda\lambda'} t}  \left(\rho_\lambda-\rho_{\lambda'}\right) |\qop_{\lambda \lambda'}|^2\right) \text{. } 
\end{align*}
All that remains to be shown is that the term in parenthesis is \(\chi(t,T)\).
To see this, it is sufficient to expand \cref{def:lresp:chi} in the eigenbasis of the equilibrium Hamiltonian \(H_0\) as
\begin{align*}
	\chi(t,T)& = i \theta(t) \braket{[\qop(t),\qop(0)]}= i\theta(t) \sum_\lambda \rho_\lambda \sum_{\lambda'}\left( e^{+i \epsilon_\lambda t}\qop_{\lambda \lambda'} e^{-i \epsilon_{\lambda'} t}\qop_{\lambda'\lambda}-\qop_{\lambda\lambda'}e^{+i \epsilon_\lambda' t}\qop_{\lambda' \lambda} e^{-i \epsilon_{\lambda} t}\right)\\
		 & = i\theta(t) \sum_{\lambda,\lambda'}e^{-i \omega_{\lambda\lambda'} t}\left(\rho_\lambda-\rho_{\lambda'}\right) |\qop_{\lambda \lambda'}|^2\text{ .}
\end{align*}

Once \cref{eq:qfi:chit} is available, it is possible to obtain expressions such as \cref{eq:qfi:quench} by applying a deconvolution procedure to the Kubo formula to extract \(\chi(t,T)\) from \(\Delta \qop(t)\) and \(f(t)\).
Importantly, the procedure works for arbitrary well-behaved quench protocols \(f(t)\).
Analytically, the deconvolution can be carried out in frequency space where the Kubo formula reads \(\Delta\widehat{\qop}(\omega) = \widehat{f}(\omega)\widehat{\chi}(\omega,T)\), which can be formally rewritten as \(\widehat{\chi}(\omega,T) =\widehat{f}(\omega)^{-1}\Delta\widehat{\qop}(\omega)\).
By applying an inverse Fourier transform, we obtain 
\begin{align}
	\chi(t,T) &= \left(\Delta \qop * v_f\right)(t)=\int_{-\infty}^{+\infty} \dv{\tau} \Delta \qop(\tau) v_f(t-\tau)
	\label{eq:lresp:deconv}
\end{align}
for \(v_f(t)=(2\pi)^{-1}\int_{-\infty}^{+\infty} \dv{\omega} e^{i\omega t} \widehat{f}(\omega)^{-1}\).
Here, \(f(t)\), \(v_f(t)\) and their Fourier transforms should be understood as generalized functions.
The inverse relation that defines \(\widehat{f}(\omega)^{-1}\) encodes that \(\phi*f*v_f=\phi\) must hold for any well behaved test function \(\phi\).
Combining \cref{eq:qfi:chit} and \cref{eq:lresp:deconv} with the fact that, due to causality, \(\Delta \qop(t)=0\) for \(t<0\), we obtain
\begin{align}
	\fq &= 4T \int_{0}^{+\infty}\dv{t} \frac{1}{\sinh\left(\pi t T\right)}\int_{0}^{\infty} \dv{\tau} \Delta \qop(\tau)v_f(t-\tau)= 4T \int_{0}^{\infty} \dv{\tau} \Delta \qop(\tau) \kappa_{f}(\tau,T)
	\label{eq:qfi:f(t)}
\end{align}
with the kernel function \(\kappa_f(\tau,T)=\int_{0}^{\infty}\dv{t} \frac{v_f(t-\tau)}{\sinh\left(\pi t T\right)}\).

\subsection{Examples of different quench protocols}

The simplest example is that of a delta-pulse, \(f(t)=\qpar \delta(t)\).
In that case, \(\widehat{f}(\omega)^{-1}=\frac{1}{\qpar}\), so \(v_\text{delta-pulse}(t)=\frac{\delta(t)}{\qpar}\) and
\begin{align*}
	\fq &= \frac{4T}{\qpar} \int_{0}^{+\infty}\dv{t} \frac{\Delta \qop_\text{delta-pulse}(t)}{\sinh\left(\pi t T\right)}
\end{align*}
which just comes from the fact that \(\chi(t,T)=\qpar^{-1}\Delta \qop_\text{delta-pulse}(t)\). 

The case presented in the main text uses a step-type quench \(f(t) = \qpar \theta(t)\) so 
\begin{align}
	\widehat{f}(\omega)^{-1} = \frac{1}{\qpar}\left(\pi\delta(\omega)+\mathcal{P}\frac{1}{i\omega} \right)^{-1} = \frac{i\omega}{\qpar} \implies v_\text{quench}(t)=\frac{\delta'(t)}{\qpar}
	\label{eq:lresp:quenchdeconv}
\end{align}
and we recover \cref{eq:qfi:quench},
\begin{align}
	\fq &= \frac{4T}{\qpar} \int_{0}^{+\infty}\dv{t} \Delta \qop_\text{quench}(t)\left(-\frac{\dv{}}{\dv{\tau}}\frac{1}{\sinh\left(\pi \tau T\right)}\right)_{\tau=t} =  \frac{4\pi T^2}{\qpar} \int_{0}^{+\infty}\dv{t} \frac{\Delta \qop_\text{quench}(t)}{\sinh\left(\pi tT\right)\tanh\left(\pi tT\right)} \text{ .}
	\label{eq:qfi:quenchderivation}
\end{align}
It is straightforward to check the validity of \cref{eq:lresp:quenchdeconv} in frequency space as \(\omega\delta(\omega)=0\) or, alternatively, in real time as
\begin{align*}
	\left(\phi * \theta * \delta' \right)(t) &=\int_{-\infty}^{+\infty} \dv{\tau_2} \int_{-\infty}^{+\infty} \dv{\tau_1}\phi(\tau_1)\theta(\tau_2-\tau_1)\delta'(t-\tau_2) = \int_{-\infty}^{+\infty} \dv{\left(t-\tau_2\right)} \int_{0}^{\tau_2} \dv{\tau_1}\phi(\tau_1)\delta'(t-\tau_2)\\ 
						 &= \left(-\frac{\dv{}}{\dv{\left(t-\tau_2\right)}}\int_{0}^{\tau_2} \dv{\tau_1}\phi(\tau_1)\right)_{t-\tau_2=0} =  \left(\frac{\dv{}}{\dv{\tau_2}}\int_{0}^{\tau_2} \dv{\tau_1}\phi(\tau_1)\right)_{\tau_2=t} = \phi(t) \text{ .}
\end{align*}

Importantly, the deconvolution procedure discussed in the previous section is only possible if \(f(t)\) is well behaved.
In particular, \(\widehat{f}(\omega)\) must only have isolated zeros in the support of \(\widehat{\chi}(\omega,T)\) since \(\widehat{f}(\omega)^{-1}\) has to be well defined as a generalized function. 
A simple drive that does not fulfill this condition is \(f(t)=\sin\left(\omega_0 t\right)\) as \(\widehat{f}(\omega)=i\pi\left(\delta(\omega+\omega_0)-\delta(\omega-\omega_0)\right)\) cannot be inverted.
From a physical standpoint, this simply highlights that the time dependent perturbation under consideration must probe all frequencies of the Kubo response function \(\widehat{\chi}(\omega,T)\). 

\subsection{Experimental considerations}

In an experimental setting, where the drive function \(f(t)\) might not have a simple functional expression and both \(f(t)\) and \(\Delta \qop(t)\) will contain some noise, one can employ a direct deconvolution procedure such as the Wiener deconvolution \cite{Wiener1964}.
This yields an approximation for \(\widehat{v}_f(\omega)\) given by
\begin{align*}
	\widehat{v}_f(\omega) = \frac{\widehat{f}(\omega)^* S_{\chi}(\omega)}{|\widehat{f}(\omega)|^2 S_{\chi}(\omega)+ S_\text{n}(\omega) }
\end{align*}
where \(S_{\chi(\text{n})}(\omega)\) denotes the mean power spectral density of \(\chi\) (the noise).

Regarding the quench scenario, there are two potential concerns that we would like to discuss as they are relevant for experimental implementation.
First, in a real experiment one does not have an ideal quench \(f(t)=\qpar\theta(t)\) but some ramp \(f(t)=\qpar r(t)\) with a smooth function \(r(t)\).
This is of no concern as long as the timescales where the ramp reaches \(r(t)\approx 1\) are much smaller than the relevant timescales for the system dynamics.
Even when that is not the case, it is possible to account for the ramp rigorously by deriving the correct \(\kappa_f\) for the specific ramp profile.
For instance, for a well defined ramp \(r_{t_0}(t)\) one has
\begin{align*}
	f(t)= \qpar r_{t_0}(t) = \qpar\begin{cases}
			0,\text{ , } t\leq 0\\
			g(t/t_0)\text{ , } 0<t<t_0\\
				1\text{ , } t_0\leq t
			\end{cases}
			\implies \kappa_f(t,T) = \kappa(t,T) * \bar{g}(t/t_0)  \text{ .}
\end{align*}
where \(\bar{g}(t)\) is a filter defined by its Fourier transform \(\bar{g}(\omega t_0)=\left(e^{-i \omega t_0} + i \omega t_0 \int_0^1 \dv{y}e^{-i y \omega t_0} g(y) \right)^{-1}\) and \(\kappa(t,T)=4\pi T^2 /\left[ \qpar\sinh\left(\pi tT\right)\tanh\left(\pi tT\right)\right]\) is the kernel of the ideal instantaneous quench with \(t_0=0\) as in \cref{eq:qfi:quenchderivation}.
For the example of a linear ramp, \(g(t/t_0)=t/t_0\), we have \(\bar{g}(\omega t_0)=i\omega t_0 /\left(e^{-i \omega t_0}-1\right)\).
 
\begin{figure}
	\includegraphics[width=\columnwidth]{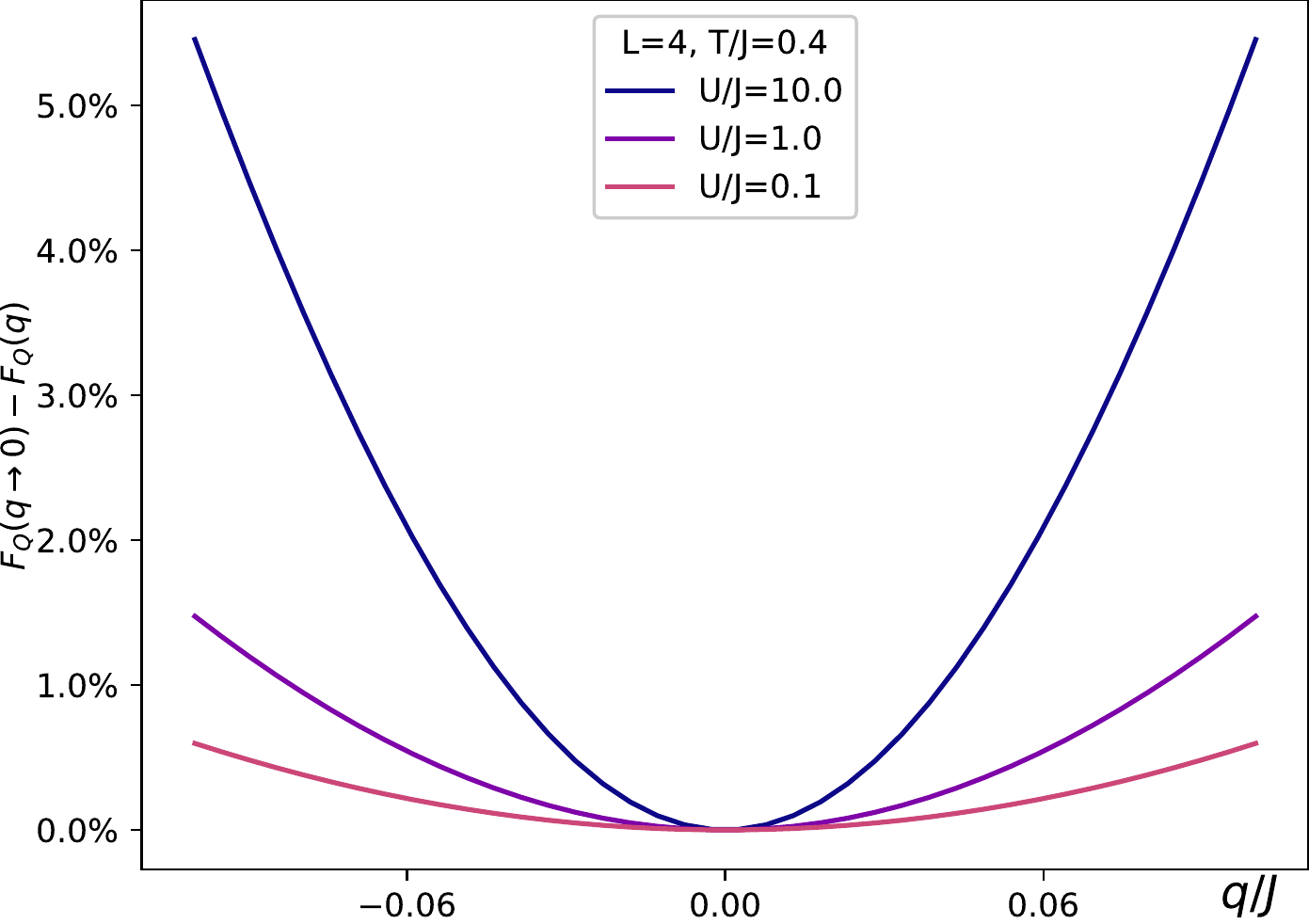}
	\caption{\textbf{QFI extraction with different \(\qpar\),} exemplified with the Fermi--Hubbard model for a quench with the staggered magnetization.
		Deviations of \(\Delta\qop_\text{quench}(t)\) from the predictions of linear response theory lead to a \(\qpar\) dependent value for \(\fq\).
		Here, the errors relative to the correct value are shown.
		In particular, a quadratic correction, coming from the \(\Delta \qop^{(3)}_\text{quench}(t)\) term, dominates at this scale so even for values of \(\qpar\) that show significant deviation from the linear response description, reasonable precision can be achieved for the QFI. 
		}
	\label{fig:quenches}
\end{figure}

The second potential concern is that, in principle, \(\Delta \qop(t)\) contains higher-order corrections, while we are only interested in the part that is described by linear response theory.
Formally, one has
\begin{align*}
	\Delta \qop_\text{quench} (t) =\sum_{n=1}^\infty \Delta \qop^{(n)}_\text{quench}(t)\qpar^n\,,
\end{align*}
while the actual term that goes into \cref{eq:qfi:quench} is \(\xi(t,T)=\Delta\qop^{(1)}_\text{quench}(t)\).
Fortunately, the linear part dominates for short times, so the exponential decay of the kernel function \(\kappa(t,T)\) mitigates any errors coming from non-linear effects.
Even more, it is possible to obtain better estimates on the linear contribution by using different values of the quench parameter \(\qpar\) and combining the results through a polynomial fit.
The simplest application relies on performing the quench with some small \(\qpar\) and with \(-\qpar\); the two measurements can then be combined to yield
\begin{align*}
	\frac{\Delta\qop_\text{quench}(t)|_{\qpar}-\Delta\qop_\text{quench}(t)|_{-\qpar}}{2} =\Delta\qop^{(1)}_\text{quench}(t)\qpar +\Delta\qop^{(3)}_\text{quench}(t)\qpar^3+\dots\,,
\end{align*}
which removes the quadratic contributions and enables one to get accurate results over larger timescales.
One can also apply this principle directly to the values of \(\fq\) as any deviations from the correct value, due to higher-order terms, will also depend algebraically on \(\qpar\).
In \cref{fig:quenches}, this is shown through the convergence of the deviations to zero as \(\qpar\) becomes infinitesimal.

\section{Entanglement Bounds}

In this section, we derive entanglement bounds for \(k\)-producible states of fixed particle number, as given in \cref{ineq:qfi:kprod}. 
For this, we start by computing the pure-state variance of the relevant operators, and find upper bounds for it through Popoviciu's inequality.
By convexity of the QFI, these yield bounds also for the case of mixed quantum states. 
Afterwards, we show how tighter bounds can be achieved in the case of fixed fermion number, as is relevant for cold-gas experiments performed at fixed atom number. 

\subsection{Detailed derivation of entanglement bounds}

Given some \(k\)-producible state \(\ket{\psi}\), as in \cref{def:state:kprodfermion}, one can compute the expectation value of the operator \(\qop\) by applying Wick's theorem,
\begin{align}
	\bra{\psi}\qop\ket{\psi}&=\sum_{m\in M} w(m) \bra{} \left(\Cop^\star_P\right)^\dagger \dots\left(\Cop^\star_2\right)^\dagger\left(\Cop^\star_1\right)^\dagger \cop^\dagger_m \cop_m  \Cop^\star_1\Cop^\star_2\dots\Cop^\star_P\ket{}=\sum_j \sum_{m\in M_j } w(m) \bra{} \left(\Cop^\star_j\right)^\dagger \cop^\dagger_m \cop_m \Cop^\star_j\ket{}\\
				&=\sum_j \sum_{m\in M_j } w(m) \sum_{\eta_j} \phi_j^\star(\eta_j) \phi_j(\eta_j) \eta_j(m) = \sum_j \sum_{\eta_j} w_j(\eta_j) p_j(\eta_j)\,.
	\label{eq:wick:qop}
\end{align}
Here, we assumed the state is normalized so that \(\bra{} \left(\Cop^\star_j\right)^\dagger \Cop^\star_j\ket{}=1\).
From the above expression, one sees that \(\bra{\psi}\qop\ket{\psi}\) is given by the sum of the expectation values of random variables \(w_j\), defined by \(w_j(\eta_j)=\sum_{m\in M_j} w(m) \eta_j(m)\), under the probability distribution \(p_j(\eta_j) =  \phi_j^\star(\eta_j) \phi_j(\eta_j)\).

Analogously, 
\begin{align}
	\bra{\psi}\qop^2\ket{\psi}&=\sum_{m\in M} \sum_{m'\in M}w(m)w(m') \bra{} \left(\Cop^\star_P\right)^\dagger \dots\left(\Cop^\star_2\right)^\dagger\left(\Cop^\star_1\right)^\dagger \cop^\dagger_m \cop_m \cop^\dagger_{m'} \cop_{m'} \Cop^\star_1\Cop^\star_2\dots\Cop^\star_P\ket{}\\
				  &=\sum_{j\neq j'} \sum_{m\in M_j} \sum_{m'\in M_{j'}}w(m)w(m') \bra{} \left(\Cop^\star_j\right)^\dagger \cop^\dagger_m \cop_m \Cop^\star_j\ket{}\bra{} \left(\Cop^\star_{j'}\right)^\dagger \cop^\dagger_{m'} \cop_{m'} \Cop^\star_{j'}\ket{}\\
				  &+\sum_{j} \sum_{m,m'\in M_j} w(m)w(m') \bra{} \left(\Cop^\star_j\right)^\dagger \cop^\dagger_m \cop_m \cop^\dagger_{m'} \cop_{m'} \Cop^\star_j\ket{}\\
				  &=\sum_{j\neq j'}\left(  \sum_{\eta_j} w_j(\eta_j) p_j(\eta_j)\right) \left(  \sum_{\eta_{j'}} w_{j'}(\eta_{j'}) p_{j'}(\eta_{j'})\right)
				  +\sum_{j} \left(  \sum_{\eta_j} w_j(\eta_j)^2 p_j(\eta_j)\right)\,,
	\label{eq:wick:qop2}
\end{align}
and we get
\begin{align}
	\var{\qop} = \bra{\psi} \qop^2\ket{\psi}-\bra{\psi} \qop\ket{\psi}^2 = \sum_j \left(  \sum_{\eta_j} w_j(\eta_j)^2 p_j(\eta_j)\right)-\left(  \sum_{\eta_j} w_j(\eta_j) p_j(\eta_j)\right)^2 = \sum_j\var{w_j}
	\label{eq:wick:var}
\end{align}
as the crossed terms \(j\neq j'\) cancel out.
Hence, we conclude that the QFI of a \(k\)-producible pure state is given by \(\fq = \sum_j 4 \var{w_j}\) as claimed in the main text.

To obtain a useful, general bound on the QFI, it is necessary to find a bound for \(\sum_j \var{w_j}\) that depends neither on the probability distributions \(p_j\) nor on the specific partitions \(M_j\), as these are state dependent, but solely on the \(k\)-producibility and the values of the \(w(m)\).
To do so, we first find a bound that assumes a given partition and then optimize the bound to find the worst-case scenario.

Let us divide each \(M_j\) into \(M_j^+=\{m\in M_j | w(m)>0\}\), \(M_j^0=\{m\in M_j | w(m)=0\}\) and \(M_j^-=\{m\in M_j | w(m)<0\}\).
Then, 
\begin{align}
	w_j(\eta_j)=\sum_{m\in M_j^+} |w(m)|\eta_j(m)-\sum_{m\in M_j^-} |w(m)|\eta_j(m) \implies  \sum_{m\in M_j^-} w(m)\leq w_j(\eta_j)\leq\sum_{m\in M_j^+} w(m)\,,
	\label{ineq:var:prepopoviciugeneral}
\end{align}
and it follows from Popoviciu's inequality that
\begin{align}
	\sum_j \var{w_j}\leq  \sum_j \frac{1}{4} \left( \sum_{m\in M_j^+} w(m) - \sum_{m\in M_j^-} w(m)\right)^2=\frac{1}{4}\sum_j \left( \sum_{m\in M_j} |w(m)|\right)^2\,,
	\label{ineq:var:popoviciugeneral}
\end{align}
where the sum over \(M_j^+\) corresponds to the maximum of \(w(\eta_j)\) over all \(\eta_j\) and the sum over \(M_j^-\) to the minimum.
The right hand side of \cref{ineq:var:popoviciugeneral} is an upper bound for the \cref{ineq:qfi:popoviciu} in the main text.

The \(k\)-partition \(\bar{M}_j\) that maximizes the right hand side of \cref{ineq:var:popoviciugeneral} can be constructed by concentrating the modes with the highest weight, in absolute value, in the same partition.
More explicitly, if we enumerate the modes \(m_1,m_2,\dots m_{|M|}\) such that \(|w(m_1)| \geq |w(m_2)| \geq \dots \geq |w(m_{|M|})|\), then
\begin{itemize}
	\item \(\bar{M}_1=\{m_1,m_2,\dots m_k\}\)
	\item \(\bar{M}_2=\{m_{k+1},m_{k+2},\dots m_{k+k}\}\)
	\item \dots
	\item \(\bar{M}_d=\{m_{(d-1)k+1},m_{(d-1)k+2},\dots m_{(d-1)k+k}\}\)
	\item \(\bar{M}_{d+1}=\{m_{dk+1},m_{dk+2},\dots m_{|M|}\}\), 
\end{itemize}
where \(|M|=dk+r\).
Hence, we can calculate specific values for a given choice of the weights to obtain the bounds,
\begin{align}
	\fq \leq \sum_{j=1}^{d+1} \left( \sum_{m\in \bar{M}_j} |w(m)|\right)^2 \leq (dk^2+r^2)\left(\max_{m\in M} |w(m)|\right)^2\,,
	\label{ineq:qfi:kprodgeneral}
\end{align}
for any \(k\)-producible pure state and it follows from the convexity of the QFI \cite{BraunsteinCaves1994,HyllusPezzeSmerzi2012} that the same holds for any \(k\)-separable mixed state.

\subsection{Tighter bounds at fixed fermion number}

The lower and upper bounds for \(w(\eta_j)\), shown in \cref{ineq:var:prepopoviciugeneral}, cannot always be reached if there are restrictions on the occupation numbers. 
For instance, if we assume that \(\ket{\psi}\sim\ket{\psi_1}\wedge\ket{\psi_2}\wedge\dots\wedge\ket{\psi_P}\) has a fixed total occupation number \(N\), i.e., \(N=\sum_{m\in M} \cop^\dagger_m \cop_m\) is a conserved quantity, then the states of individual partitions, \(\ket{\psi_j}=\Cop^\star_j\ket{}\), must also have fixed occupation numbers \(N_j\) such that \(\sum N_j =N\).
If that is the case, then the number of modes in \(\bar{M}^+_j\) and \(\bar{M}^-_j\) has to be the same and equal to \(N_j\) if the \(\eta_j\)'s should be capable of achieving the limits of \cref{ineq:var:prepopoviciugeneral}.
While the bound \cref{ineq:qfi:kprodgeneral} still holds, it is possible to derive tighter bounds by exploiting these facts. 

To obtain the improved bounds, let us again consider an arbitrary partition and divide each \(M_j\) into \(M_j =M_j^l \cup M_j^i \cup M_j^u\) where the lower portion \(M_j^l\) contains the \(N_j\) modes with the lowest weights, the upper portion \(M_j^u\) contains the \(N_j\) modes with the highest weights, and the intermediary portion \(M_j^i\) contains the rest of the modes.
It follows that, for any \(\eta_j\) that respects the constraint of having occupation number \(N_j\), 
\begin{align}
	\sum_{m\in M_j^l} w(m)\leq w_j(\eta_j)\leq\sum_{m\in M_j^u} w(m)\,.
	\label{ineq:var:prepopoviciufixed}
\end{align}
We can apply again Popoviciu's inequality to obtain
\begin{align}
	\sum_j \var{w_j}\leq  \sum_j \frac{1}{4} \left( \sum_{m\in M_j^u} w(m) - \sum_{m\in M_j^l} w(m)\right)^2\,.
	\label{ineq:var:popoviciufixed}
\end{align}
The task is now to find the \(k\)-partition \(\bar{\bar{M}}_j\) that optimizes the right hand side of \cref{ineq:var:popoviciufixed}.
This can be done by concentrating as much as possible the modes with the highest weights into the same \(M_j^u\)'s and those with the lowest weights into the same \(M_j^l\)'s in a similar fashion as above.
If we enumerate the modes \(m_1,m_2,\dots m_{|M|}\) such that \(w(m_1) \geq w(m_2) \geq \dots \geq w(m_{|M|})\), then
\begin{itemize}
	\item \(\bar{\bar{M}}_1=\{m_1,m_2,\dots m_{k/2}\}\cup\{m_{|M|}, m_{|M|-1}, \dots m_{|M|-k/2+1}\}\)
	\item \(\bar{\bar{M}}_2=\{m_{k/2+1},m_{k/2+2},\dots m_{k/2+k/2}\}\cup\{m_{|M|-k/2},m_{|M|-k/2-1}, \dots m_{|M|-k/2-k/2+1}\}\)
	\item \dots
	\item \(\bar{\bar{M}}_d=\{m_{(d-1)k/2+1},m_{(d-1)k/2+2},\dots m_{(d-1)k/2+k/2}\}\cup\{m_{|M|-(d-1)k/2},m_{|M|-(d-1)k/2-1}, \dots m_{|M|-(d-1)k/2-k/2+1}\}\)
	\item \(\bar{\bar{M}}_{d+1}=\{m_{dk/2+1},m_{dk/2+2},\dots m_{|M|-dk/2}\}\)\,.
\end{itemize}
Here, for simplicity we assume \(k\) is even, but it is straightforward to correct the indexes when \(k\) is odd.
The above partition already takes into account the fact that we also have to optimize over the numbers \(N_j\) as the bound can only depend on the total number \(N\).
The algorithm below describes how to allocate the optimal choice \(\bar{\bar{N}}_j\).
It essentially consists in keeping as many partitions at half-filling as possible, with priority given to the initial ones as they contribute more to \cref{ineq:var:popoviciufixed}:
\begin{enumerate}
	\item Let \(\tilde{N}\) denote the number of unallocated fermions. Initially \(\tilde{N}=N\)
	\item Initialize all \(\bar{\bar{N}}_1,\bar{\bar{N}}_2,\dots \bar{\bar{N}}_{d+1}\) to zero
	\item For \(j=1,\dots d+1\) do:
		\begin{itemize}
			\item if \(\tilde{N}\geq |\bar{\bar{M}}_j|/2\) then \(\bar{\bar{N}}_j = \bar{\bar{N}}_j + |\bar{\bar{M}}_j|/2\) and \(\tilde{N}= \tilde{N}-|\bar{\bar{M}}_j|/2\)
			\item if \(\tilde{N}< |\bar{\bar{M}}_j|/2\) then \(\bar{\bar{N}}_j = \bar{\bar{N}}_j + \tilde{N}\) and \(\tilde{N}= 0\)
		\end{itemize}
	\item If \(\tilde{N}\neq 0\) then for \(j=d+1,\dots 1\):
		\begin{itemize}
			\item if \(\tilde{N}\geq |\bar{\bar{M}}_j|/2\) then \(\bar{\bar{N}}_j = \bar{\bar{N}}_j + |\bar{\bar{M}}_j|/2\) and \(\tilde{N}= \tilde{N}-|\bar{\bar{M}}_j|/2\)
			\item if \(\tilde{N}< |\bar{\bar{M}}_j|/2\) then \(\bar{\bar{N}}_j = \bar{\bar{N}}_j + \tilde{N}\) and \(\tilde{N}= 0\)
		\end{itemize}
\end{enumerate}

Combining the two ingredients, the optimal partition and occupations, it now becomes possible to get specific bounds for different situations.
In particular, we get a generic bound
\begin{align}
	\fq \leq \frac{dk^2+r^2}{4}\left(\max_{m\in M} w(m)-\min_{m\in M} w(m)\right)^2
	\label{ineq:qfi:kprodfixed}
\end{align}
for any state with a fixed occupation number.
Bounds for specific fillings will be tighter and can be obtained by explicitly calculating the right hand side of \cref{ineq:var:popoviciufixed} using the optimal partition and occupations just described.
Notably, a similar reasoning can be applied to the bosonic particles with fixed total occupation number as well as to the usual bounds for spin systems if the polarization is fixed.

\end{document}